\documentclass[twoside]{epws08}
\usepackage[latin1]{inputenc}
\usepackage[dvips]{graphicx,epsfig,color}
\usepackage{wrapfig,rotating}

\usepackage{amsmath}                    
\usepackage{amssymb}                    
\usepackage{amstext}                    
\usepackage{amsfonts}                   
\usepackage{array}                      
\usepackage{latexsym}                   
\usepackage{wasysym}                    %

\newcommand{\ALR}    {\ensuremath{ A_{\mathrm{LR}}} }
\newcommand{\ppl}    {\ensuremath{{\mathcal P}_{e^+}} }
\newcommand{\pmi}    {\ensuremath{{\mathcal P}_{e^-}} }

\newcommand{\peff}   {\ensuremath{{\mathcal P}_{\rm{eff}}} }

\definecolor{lgrey}      {rgb}{0.80, 0.80, 0.80}  
\definecolor{mgrey}      {rgb}{0.52, 0.52, 0.52}  
\definecolor{dgrey}      {rgb}{0.28, 0.28, 0.28}  
\definecolor{rred}       {rgb}{1.00, 0.00, 0.00}  
\definecolor{lred}       {rgb}{1.00, 0.65, 0.70}  
\definecolor{mred}       {rgb}{1.00, 0.10, 0.10}  
\definecolor{dred}       {rgb}{0.70, 0.00, 0.00}  \newcommand{\dred}       {\color{dred}}
\definecolor{ddred}      {rgb}{0.55, 0.00, 0.00}  
\definecolor{burntorange}{rgb}{0.70, 0.20, 0.10}  
\definecolor{lorange}    {rgb}{1.00, 0.85, 0.00}  
\definecolor{orange}     {rgb}{1.00, 0.85, 0.30}  
\definecolor{ggreen}     {rgb}{0.05, 1.00, 0.10}  
\definecolor{dgreen}     {rgb}{0.02, 0.40, 0.05}  
\definecolor{mgreen}     {rgb}{0.10, 1.00, 0.10}  
\definecolor{greygreen}  {rgb}{0.18, 0.40, 0.30}  
\definecolor{greenish}   {rgb}{0.00, 0.68, 0.58}  
\definecolor{neongreen}  {rgb}{0.00, 0.80, 0.20}  
\definecolor{bblue}      {rgb}{0.10, 0.10, 0.95}  
\definecolor{lblue}      {rgb}{0.78, 0.88, 1.00}  
\definecolor{mblue}      {rgb}{0.20, 0.36, 0.72}  
\definecolor{dblue}      {rgb}{0.02, 0.05, 0.54}  \newcommand{\dblue}      {\color{dblue}}
\definecolor{blueish}    {rgb}{0.25, 0.28, 0.78}  
\definecolor{midnight}   {rgb}{0.00, 0.45, 0.50}  
\definecolor{orchid}     {rgb}{0.74, 0.50, 0.76}  
\definecolor{violet}     {rgb}{0.70, 0.00, 0.70}  
\definecolor{dviolet}    {rgb}{0.47, 0.00, 0.45}  
\definecolor{gviol}      {rgb}{0.56, 0.54, 0.80}  
\definecolor{dgviol}     {rgb}{0.46, 0.44, 0.68}  
\definecolor{lrose}      {rgb}{0.84, 0.80, 1.00}  
\definecolor{lwhite}     {rgb}{1.00, 1.00, 1.00}  
\definecolor{dblack}     {rgb}{0.00, 0.00, 0.00}  
\definecolor{lcyan}      {cmyk}{1.00, 0.00, 0.00, 0.05}  
\definecolor{dcyan}      {cmyk}{0.70, 0.00, 0.00, 0.15}  
\definecolor{beige}      {cmyk}{0.05, 0.20, 0.90, 0.15}  

\begin{document}
\title{
  \vspace*{-10mm}
  \begin{flushright}
    {\normalsize\textnormal
      {\sc
        ILC-NOTE-2008-047 \\
        DESY-08-099    \\
        SLAC-PUB-13296 \\[2mm]
      }
      {\textnormal \today} \\[12mm]
    }
  \end{flushright}
  Executive Summary of the Workshop on Polarisation and Beam Energy Measurements at the ILC \\[2mm]
} 
\author{
  B.~Aurand$^{1}$,         
  I.~Bailey$^{2}$,     	   
  C.~Bartels$^{3}$,        
  G.~Blair$^{4}$,          
  A.~Brachmann$^{5}$,      
  J.~Clarke$^{6}$,         
  L.~Deacon$^{4}$, \\      
  V.~Duginov$^{7}$,        
  A.~Ghalumyan$^{8}$,      
  A.~Hartin$^{3,9}$,       
  J.~Hauptman$^{10}$,      
  C.~Helebrant$^{3}$,      
  S.~Hesselbach$^{11}$, \\ 
  D.~K\"afer$^{3}$\footnote{Corresponding author: daniela.kaefer@desy.de} , 
  J.~List$^{3}$,           
  W.~Lorenzon$^{12}$,      
  A.~Lyapin$^{13}$,        
  I.~Marchesini$^{3}$,     
  R.~Melikian$^{8}$,       
  K.~M\"onig$^{3}$, \\     
  K.C.~Moffeit$^{5}$,      
  G.~Moortgat-Pick$^{11}$, 
  N.~Muchnoi$^{14}$,       
  V.~Nikoghosyan$^{8}$,    
  S.~Riemann$^{3}$, \\     
  A.~Sailer$^{3}$,         
  A.~Sch\"alicke$^{3}$,    
  H.J.~Schreiber$^{3}$,    
  P.~Sch\"uler$^{3}$,      
  P.~Starovoitov$^{15}$,   
  E.~Torrence$^{16}$, \\   
  A.~Ushakov$^{3}$,        
  U.~Velte$^{3}$,          
  M.~Viti$^{3}$,           
  G.~Weiglein$^{11}$,      
  J.~Wittschen$^{1}$,      
  M.~Woods$^{5}$           
  \vspace{4mm} \\
  1- Phys. Inst., University of Bonn, Germany;\quad
  2- Cockcroft Inst., University of Liverpool, UK; \\
  3- DESY, Hamburg and Zeuthen, Germany;\quad 
  4- Royal Holloway, University of London, UK; \\
  5- SLAC, Stanford, USA;\quad 
  6- Daresbury Laboratory, UK;\quad 
  7- JINR, Dubna, Russia; \\ 
  8- Yerevan Physics Institute, Armenia;\quad 
  9- JAI, Oxford, UK;\quad  
  10- Iowa State University, USA; \\
  11- IPPP, University of Durham, UK;\quad 
  12- University of Michigan, USA; \\
  13- University College London, UK;\quad 
  14- Budker INP, Novosibirsk, Russia; \\
  15- NCPHEP, Minsk, Belarus;\quad 
  16- University of Oregon, USA
}

\maketitle

\vspace*{2mm}
\begin{abstract}
This note summarizes the results of the ``Workshop on Polarisation and 
Beam Energy Measurements at the ILC'', held at DESY (Zeuthen) April 9-11 2008. 
The topics for the workshop included 
  (i) physics requirements, 
 (ii) polarised sources and low energy polarimetry, 
(iii) BDS polarimeters, 
 (iv) BDS energy spectrometers, and 
  (v) physics-based measurements of beam polarisation 
      and beam energy from collider data.
Discussions focused on the current ILC baseline programme as described in the 
Reference Design Report (RDR), which includes physics runs at beam energies 
between 100 and 250~GeV, as well as calibration runs on the $Z$-pole. 
Electron polarisation of $\pmi \apprge 80\,$\% and 
positron polarisation of $\ppl \apprge 30\,$\% are part of the baseline 
configuration of the machine. Energy and polarisation measurements for 
ILC options beyond the baseline, including $Z$-pole running and the 1~TeV 
energy upgrade, were also discussed. \\
\end{abstract}

\newpage
{\bf \noindent
  Seven recommendations requiring follow-up from the GDE and the Research 
  Director have emerged from the workshop: 
  \renewcommand{\labelenumi}{\arabic{enumi}. }
  \begin{enumerate}
  \item  Separate the functions of the upstream polarimeter chicane. 
    Do not include an MPS energy collimator or laser-wire emittance 
    diagnostics; use instead a separate 
    setup for these two. 
  \item  Modify the extraction line polarimeter chicane from a 4-magnet 
    chicane to a 6-magnet chicane to allow the Compton electrons to be 
    deflected further from the disrupted beam line.
  \item  Include precise polarisation and beam energy measurements 
    for {\boldmath$Z$}-pole calibration runs into the baseline configuration.
  \item  Keep the initial positron polarisation of 30-45\% for physics (baseline).
  \item  Implement parallel spin rotator beamlines with a kicker system before 
    the damping ring to provide rapid helicity flipping of the positron spin. 
  \item  Move the pre-DR positron spin rotator system from 5~GeV to 400~MeV. 
    This eliminates expensive superconducting magnets and reduces costs.
  \item  Move the pre-DR electron spin rotator system to the source area. 
    This eliminates expensive superconducting magnets and reduces costs.
  \end{enumerate}
}

\vspace*{2mm}
\section{Introduction}
With the foreseen start of the Large Hadron Collider (LHC) in 2008, direct 
discoveries of new particles beyond the Standard Model (SM) are expected soon. 
According to the currently envisioned timeline for the International Linear 
Collider (ILC), first physics results could be available in 2020. 
Compared to the new energy frontier opened by the LHC, the ILC will open 
a new precision frontier, with beam polarisation playing a key role in 
a physics programme that demands precise polarisation and beam energy 
measurements~\cite{bib:RDRphys}.

In compliance with the RDR~\cite{bib:RDRacc}, the baseline configuration 
of the ILC already provides polarised electron and positron beams, 
with spin rotator systems to achieve longitudinal polarisation at 
the collider-IP; 
upstream and downstream polarimeters and energy spectrometers for 
both beams; and the capability to rapidly flip the electron helicity 
at the injector (using the source laser). 
Only the possibility of fast positron helicity flipping is not 
included in the baseline configuration.

The electrons will be highly polarised ($\pmi \apprge 80\,$\%), 
but also the positrons will be produced with an initial 
polarisation of about 30-45\%. This expected small positron 
polarisation can either be destroyed or it can be used with 
great benefit for physics measurements if the possibility of 
fast helicity flipping of the positron spin is also provided. 
Excellent polarimetry accurate to $\Delta \mathcal{P}/\mathcal{P} = 0.25$\% 
has been assumed in reference~\cite{bib:POWER}. 
This number is technology-driven and can only be achieved with 
dedicated Compton polarimeters located upstream and downstream 
of the $e^+e^-$ interaction region. 

Precise beam energy measurements are necessary at the ILC in order to 
measure particle masses produced in high-rate processes. Measuring the 
top mass in a threshold scan to order 100~MeV or measuring a Standard 
Model Higgs mass in direct reconstruction to order 50~MeV~\cite{bib:RDRphys} 
requires knowledge of the luminosity-weighted mean collision energy 
$\langle \sqrt{s} \rangle$ to a level of $1 - 2 \cdot 10^{-4}$. 
Precise measurements of the incoming beam energy are a critical component 
to measuring the quantity $\langle \sqrt{s} \rangle$ as it sets the overall 
energy scale of the collision process. 

Although precision polarisation and energy measurements at the $Z$-pole are not part 
of the current baseline as described in the RDR~\cite{bib:RDRacc}, we argue that 
the baseline should be modified to include such measurements. $Z$-pole calibration 
data would provide a unique possibility for an early calibration of the polarimeters 
and energy spectrometers in a well understood physics regime. Additionally, the 
calibration data can provide precision measurements of electroweak observables and 
thus serve as extremely sensitive tests of the SM, if the beam polarisation and 
energy are accurately measured. 
This is discussed in detail in a separate paper~\cite{bib:Zpole-Gudi}.

In any case, the actual polarisation state -- as well as the energy of the 
beam -- has to be known precisely to ensure the foreseen high precision physics 
measurements. The machine parameters needed for a linear collider to fulfill these 
physics requirements have been worked out~\cite{bib:scope}. It is mandatory that 
these requirements on polarisation, beam energy and luminosity measurements are 
technically achievable. 
Comparing the above quoted numbers with the typical precisions aimed for in the ILC 
physics program, it is clear that the goal for polarisation and energy measurements 
is limited by technology, whereas the physics programme would benefit from an even 
higher precision of both measurements.

\vspace*{1mm}
\section{Polarimetry} \label{sec:polarimetry}
\subsection{Sources,  Low Energy Polarimetry and Spin Rotation}
The electron source produces polarised electrons from a DC photocathode gun. 
The circular polarisation of the source laser beam is set with fast Pockels 
cells and the laser helicity can be reversed train-to-train, thereby allowing 
fast reversals of the electron spin. A Mott polarimeter located in a diagnostic 
line will be used to determine the electron polarisation near the source.

The positron source uses photoproduction to generate positrons~\cite{bib:RDRacc}. 
The electron main linac beam passes through a long helical undulator 
to generate a multi-MeV photon beam, which then strikes a thin metal 
target to generate positrons in an electromagnetic shower. 
The positrons are captured, accelerated, separated from the shower 
constituents and unused photon beam and then are transported to the 
Damping Ring. Although the baseline design only requires unpolarised 
positrons, the positron beam produced by the baseline source has a 
polarisation of  $\ppl \apprge 30\,$\%, and beamline space has been 
reserved for an eventual upgrade to 60\% polarisation.
No low energy polarimetry for the positron beam is foreseen in the RDR, 
but R\&D work is ongoing. The positron polarisation could be measured 
near the source after the pre-accelerator using a Bhabha polarimeter 
at 400~MeV. After the damping ring, the positron polarisation could be 
checked with a Compton polarimeter. To save costs the laser of the laser 
wire system could be used, but design studies are not yet done.
Since the RDR was written, simulation studies show that bunch 
(energy) compression would increase the positron capture effiency 
at the source, with which the positron polarisation could even 
reach $\ppl \apprge 45\,$\% at the beginning of the ILC physics 
program.~\cite{bib:e+pol45}.

There are two ways to use the positron beam: 
 (i) Physics measurements with a positron polarisation 
     of about $\ppl \apprge 30-45\,$\%. 
(ii) Unpolarised positrons at the $e^+e^-$--IP. 
In the first case (i), the polarised positron beam is transported to the 
$e^+e^-$--IP with minimal spin diffusion and the degree of polarisation 
is measured with high precision of 0.25\% near the interaction region 
with upstream and downstream polarimeters. 
Appropriate spin rotator systems are described in the RDR and are 
included in the beam transport lines from the Linac to the Damping 
Ring (LTR) and from the Damping Ring to the Linac (RTL) for both 
electrons and positrons.
The left-right asymmetric structure of Standard Model processes 
requires a particular configuration of the initial state helicities, 
which should be randomly available to minimize systematic effects. 
So the effective luminosity can be increased, e.g. by a factor of~1.24 
compared to $\pmi \apprge 80\,$\% and zero positron polarisation, 
and the uncertainty of the effective polarisation given by 
$\peff = \frac{|\pmi|+|\ppl|}{1+|\pmi|\,|\ppl|}$ 
can be reduced. The actual frequency of the helicity flip depends 
on the long time stability and reproducibility of machine parameters 
as luminosity, polarisation and background conditions. A helicity 
reversal for positrons less frequent than for electrons will cancel 
the gain in effective luminosity and would also reduce the improvement 
for the polarisation uncertainty. 
In the current baseline design, however, the positron helicity can 
only be slowly reversed by changing the polarity of the superconducting 
spin rotator magnets. This does not satisfy the demands of the precision 
physics program, which needs positron helicity reversals train-to-train 
as it is done for electrons. \\
{\it\dred Recommendation:
  Keep the initial positron polarisation of 30-45\% (baseline).
  Modify the baseline configuration to provide random selection 
  of the positron helicity train-by-train by implementing parallel 
  spin rotator beamlines and kicker systems in the positron 
  LTR~\cite{bib:SpinRotSchemes-2005}.
}

Positron spin rotation and flipping could be done at 400 MeV rather than 
at 5~GeV~\cite{bib:SpinRotLowE-2008}, while the electron spin rotation 
could be done at the electron source using a Wien filter. 
The solenoid magnets necessary to rotate the spin from the transverse 
horizontal to the vertical direction can be smaller and less expensive, 
demanding less tunnel space at 400~MeV compared to 5~GeV. These 
modifications would eliminate expensive superconducting magnets, 
simplify the engineering for these systems, and reduce the costs. \\
{\it\dred Recommendation:
  Move the pre-damping-ring spin rotator systems to lower energy 
  for both beams, electrons and positrons~\cite{bib:SpinRotLowE-2008}.
}

If, in the second case (ii), it should be decided to not deliver 
the 30-45\% positron polarisation from the source to the experiment, 
a special scheme after the positron damping ring needs to be devised 
to completely destroy the positron polarisation in order not to 
adversely effect the physics measurements\footnote{Spin 
  tracking studies~\cite{bib:Larisa-ST} have shown that the 
  horizontal projections of the spin vectors of an $e^+$ or 
  $e^-$ bunch do not fully decohere in the damping ring, 
  even after 8000 turns.}.
The zero positron polarisation also needs to be measured with high 
precision of 0.25\%. Further studies are needed to ensure a left-over 
positron DC polarisation of about 0.1\% will not affect physics 
measurements, which could result in the need for an even higher 
precision in this case. 

In both cases it is required to measure the positron polarisation 
with high precision. We strongly recommend option (i) whereby physics 
measurements are possible with a positron polarisation of 
$\ppl \apprge 30\,$\%.

\subsection{Overall Polarimetry Scheme} \label{sub:OverallPolScheme}
The ILC offers three methods to measure polarisation after acceleration: 
upstream and downstream of the IP, as well as using annihilation events. 
For the discussion on polarimetry it is important to distinguish the cases 
with and without positron polarisation. 
%
%
%
Without positron polarisation the cross section for all processes can be 
written as $\sigma = \sigma_0[1 - \pmi\,\ALR]$. In this case the error on 
$\ALR$ due to polarisation is $\Delta \ALR/\ALR = \Delta \pmi/\pmi$ but 
only the luminosity weighted averaged polarisation matters. If also positron 
polarisation is available the cross section can typically be written as 
$\sigma = \sigma_0 \left[1 - \ppl\,\pmi +(\ppl - \pmi)\,\ALR \right]$. 
In this case the polarisations enter linearly and as a product so that the 
correlation between the two polarisations matters. 
The relevant quantity for physics analyses in this case is an effective 
polarisation, e.g.~$\peff = \frac{|\pmi|+|\ppl|}{1+|\pmi|\,|\ppl|}$, 
which due to favourable error propagation reduces the polarisation 
uncertainty by a factor of up to three.

Apart from the polarimeters, polarisation can also be measured using annihilation 
data~\cite{bib:PolScheme-Moenig}, with direct access to the luminosity weighted 
polarisation. With electron and positron polarisation, four polarisation combinations 
are measurable for four unknowns so that the polarisation can be measured in a 
model independent way. With only electron polarisation this is not possible. 
However, W-pair production can be used to determine electron polarisation when 
only one beam is polarised with the only assumption that the $e\nu W$-coupling 
is purely left-handed which is well tested. 
The forward peak is entirely dominated by t-channel neutrino exchange and not 
influenced by possibly unknown triple gauge interactions. In both cases a 0.1\% 
precision on the individual polarisations is possible, where, due to the favourable 
error correlation, the effective polarisation can even be measured to the 0.02\% level.  
Nevertheless, annihilation methods can only provide polarisation measurements 
on very long time scales ($\apprge$ months) and need corrections from the polarimeters. 
Also, the model independent scheme with positron polarisation needs some statistics 
on all four helicity combinations, i.e. approximately about 30\% of the running 
time~\cite{bib:Zpole-Gudi, bib:ECFA08-WarsawTalk-Moenig} has to be spent on the 
less interesting J=0 combinations. If the only reason to run at these states is 
polarimetry, polarisation measurements from annihilation data are fairly expensive. 

The two polarimeters are highly complementary. The downstream polarimeter has 
access to the depolarisation in the interaction while the upstream polarimeter 
has a much higher counting rate and time granularity which is important for 
correlation measurements. Both properties are needed for a high precision analysis. 
Outside collisions the two polarimeters can calibrate each other.

To obtain a useful polarisation measurement the beam trajectories are required to be 
aligned to less than 50~$\mu$rad at the upstream Compton-IP, the collider-IP, and the 
downstream Compton-IP. This should be achievable by the beam delivery system (BDS) 
alignment as described in the RDR. 
However, the impact of the IR magnets and the crossing angle on the spin alignment 
needs to be addressed more thoroughly. In the extraction line, corrector magnets 
are needed to successfully compensate possible deflections resulting from misaligned 
beam and detector solenoid axes. 
The upstream polarimeter system, which is about 1800~m upstream of the $e^+e^-$--IP 
with a 1.5~m horizontal offset will require precision alignment methods. 
In addition, it should be possible to correlate polarimeter measurements with 
local BPM measurements, and the downstream polarimeter will want to correlate 
its measurements with the BPM measurements at the $e^+e^-$--IP~\cite{bib:EPWS-Woods-MDIissues}. 
This requires the BPM system to provide information to the 
polarimeter DAQ including bunch number identification. 
(Toroid information will also need to be provided to the DAQ.) 

For the final polarisation measurement at the ILC it is therefore indispensable 
to have upstream and downstream polarimetry and get an absolute calibration from 
annihilation data. The polarimeters provide corrections and measure the polarisation 
on short scales like for individual scan points in an energy scan, while the 
annihilation data can check the absolute calibration on very long timescales. 
To keep the corrections small, every effort should be made to flip electron and 
positron polarisation frequently, if possible train by train. For the small errors 
envisaged at the ILC, a possible cross check of the different ways to measure 
polarisation is mandatory. This has also been confirmed by the polarimetry 
experience at SLC and by the beam-energy measurements at both, LEP and SLC.

\subsection{The Upstream Polarimeter}
The upstream polarimeter is located at the beginning of the BDS, upstream 
of the tuneup dump and at a distance of roughly 1.8~km to the $e^+e^-$--IP. 
In this position it benefits from clean beam conditions and very low 
backgrounds compared to any location downstream of the IP. It is therefore 
suited to provide very fast and precise measurements of the polarisation 
before collisions. 

A complete conceptual layout for the upstream polarimeter had already 
been worked out for TESLA in 2001. However, for the ILC, a dedicated 
chicane-based spectrometer was adopted for upstream polarimetry in 2005, 
as this configuration allows the usage of a single laser wavelength at 
all beam energies when the spectrometer is operated with a fixed magnetic 
field. In this original design with a dedicated fixed-field chicane, the 
upstream polarimeter promised to be a superb and robust instrument with 
broad spectral coverage, very low background, excellent statistical 
performance for all machine bunches, and a high degree of redundancy. 
If equipped with a suitable laser, for example a similar one as used
at the TTF/Flash source, which is now in operation at DESY, it can
include every single bunch in the measurement. This will permit
virtually instant recognition of variations within each bunch train
as well as time dependent effects that vary train-by-train.
The statistical precision of the polarisation measurement will be
already 3\% for any two bunches with opposite helicity, which leads
to an average precision of 1\% for each bunch position in the train 
after the passage of only 20 trains (4 seconds). 
The average over two entire trains with opposite helicity will have 
a statistical error of $\Delta \mathcal{P}/\mathcal{P} = 0.1$\%.

The statistical power of the upstream polarimeter depends almost exclusively on the 
employed laser and therefore to first order factorizes from other design aspects. 
However the crucial issue which drives the design of the whole polarimeter is to 
reach an unprecedented low systematic uncertainty of $\delta P /P \leq 0.25\%$ or 
better~\cite{bib:EPWS-List-Upstream} with the largest uncertainties coming from 
the analyzing power calibration (0.2\%) and the detector linearity (0.1\%). \\

In an effort to reduce the cost of the long and expensive BDS system, 
the BDS management decided in autumn 2006 to combine other diagnostic 
and machine functions with the upstream polarimeter chicane. A machine 
protection system (MPS) energy collimator, defining the energy acceptance 
of the subsequent tune-up dump and a photon detector for laser-wire based 
emittance diagnostics were incorporated in the original upstream polarimeter 
chicane. The implications of these functional unification measures for 
polarimetry are rather severe and have since been the subject of protracted 
debate between the diagnostics groups and the BDS management. At this time, 
the conflicts have not yet been resolved. The following principal issues exist: 
\renewcommand{\labelenumi}{(\roman{enumi})}
\begin{enumerate}
\item {\it\dblue MPS energy collimator:}  
  The collimator is planned to be 3~m long with a $\pm 10$\% momentum aperture, 
  although there is no concrete design available yet. Its insertion into the 
  polarimeter magnetic chicane will completely obstruct the tapered vacuum 
  chamber which had been designed to minimize wakefield effects.
\item {\it\dblue Collimator generated backgrounds:}  
  Depending on the details of the structure, background generated from beam 
  halo and jitter can grow to such a degree that it becomes very difficult 
  (if not impossible) to provide a precise polarisation measurement. 
\item {\it\dblue Fixed versus scaled field operation:}  
  Fixed-field operation is the raison d'\^{e}tre of the entire chicane for 
  polarimetry, but since the MPS insertion would demand complicated and 
  costly movable jaw engineering in vacuum, the BDS management has asked 
  for a scaled-field operation. 
  While an adequate scaled-field operation, over limited energy ranges, would 
  be possible for polarimetry, the operation would be much more complicated 
  and the overall performance greatly reduced. Most importantly, it would 
  effectively render all low-energy polarimetry impossible with no prospect 
  of regaining this loss as long as the MPS object remains in this place. 
\item {\it\dblue Incorporation of emittance diagnostics:}  
  From the outset it was clear that a detector placed directly in the neutral 
  beamline would not have adequate clearance from the charged beam path in the 
  chicane at beam energies much higher than 250~GeV. 
  A detector at this location would be bombarded with synchrotron 
  radiation~\cite{bib:SRnote-MoffeitWoods} and high-energy bremsstrahlung 
  generated by beam gas interactions in the upstream beam line.
  In recognition of these problems, the laser-wire group is now exploring 
  alternatives, including indirect photon detection from a converter 
  target~\cite{bib:EPWS-Deacon-Laserwire} and Compton electron detection. 
  However, any material (e.g. converter) inserted into the neutral beam line, 
  will naturally generate more background in the polarimeter hodoscope detector, 
  thereby compromising the otherwise clean environment of the upstream 
  polarimeter.

  Compton electron detection seems to be a viable and promising alternative 
  without introducing new backgrounds. It would require the insertion of 
  retractable detectors in the chicane vacuum chamber, thus requiring some 
  nontrivial engineering. For an adequate separation of the Compton recoil 
  electrons from the original beam at low beam energies, this technique is 
  only practical for fixed-field operation of the chicane. This is just one 
  more good reason to dismiss the scaled-field scenario. 
\end{enumerate}

\noindent 
The description of the upstream polarimeter chicane combined with the MPS 
energy collimator and the laser-wire detection system given in the RDR is 
not satisfactory. The laser-wire detection system needs significant R\&D 
to demonstrate a viable system, even in a standalone system separate from 
the polarimeter chicane.

In our judgement, it has been a very unreasonable decision to place the 
MPS energy collimator into the polarimeter chicane. Apart from a host of 
very serious engineering issues, the negative impact of scaled-field operation 
is severe, particularly at low beam energies. While physics data taking at 
the $Z$-pole is not part of the ILC baseline program, the capability for 
excellent polarimetry at the $Z$-pole should not be precluded. Consequently, 
an alternative placement for the MPS collimator should be created, preferably 
in conjunction with the laser-wire emittance diagnostics. If such an alternate 
place does not exist within the baseline BDS, it will also not exist in a 
post baseline upgrade of the BDS, thereby jeopardizing $Z$-pole polarimetry. \\
{\it\dred Recommendation:
  Separate the functions of the upstream polarimeter chicane. 
  Do not include laser-wire emittance diagnostics or an MPS energy 
  collimator; use instead a separate 
  setup for these two.
}

\subsection{The Downstream Polarimeter}
The downstream polarimeter is located about 150~m downstream of the $e^+e^-$--IP 
in the extraction line and on axis with the IP and IR magnets. It can measure the 
beam polarisation both with and without collisions, thereby testing the calculated 
depolarisation correction which is expected to be at the 0.1-0.2\% level. 

A complete conceptual layout for the downstream polarimeter exists, including 
magnets, laser system and detector configuration. Three 10~Hz laser systems can 
achieve Compton collisions for three out of 2800 bunches in a train. 
Each laser will sample one particular bunch in a train for a time interval of 
a few seconds to a minute, then select a new bunch for the next time interval, 
and so on in a pre-determined pattern. 
The Compton statistics are high with about 300 Compton scattered electrons 
per bunch in a detector channel at the Compton edge. 

With this design, a statistical uncertainty of less than 1\% per minute can be 
achieved for each of the measured bunches. This is dominated by fluctuations in 
Compton luminosity due to beam jitter and laser targeting jitter and to possible 
background fluctuations. The statistical error due to Compton statistics in 
one minute, for a bunch sampled by one laser, is 0.3\%.
However, if compared to the average precision of the upstream polarimeter, 
a similar precision for each bunch position in a train could only be reached 
after about 17 hours.

Background studies have been carried out for disrupted beam losses and for the 
influence of synchrotron radiation. There are no significant beam losses for the 
nominal ILC parameter set and beam losses look acceptable even for the low power 
option. A synchrotron radiation collimator protects the Compton detector and no 
significant SR backgrounds are expected.
The systematic precision is expected to be about 0.25\%, with the largest 
uncertainties coming from the analyzing power calibration (0.2\%) and 
detector linearity (0.1\%). \\

The extraction line polarimeter chicane described in the RDR has four magnets 
with the same deflection in each magnet system. A proposal to modify the 
downstream polarimeter chicane to a six-magnet chicane was presented to the 
ILC in March 2007~\cite{bib:Moffeit-6magnets}. 
The additional two magnets after the Compton detector allow the third and 
fourth magnets of the polarimeter chicane to be operated at higher field to 
deflect the Compton electrons further from the beam line and return the beam 
to the nominal trajectory. \\
{\it\dred Recommendation:
  Modify the extraction line polarimeter chicane from a 4-magnet chicane 
  to a 6-magnet chicane to allow the Compton electrons to be deflected 
  further from the disrupted beam line.
}

\section{Beam Energy Measurements} \label{sec:BeamEnergyMeas}
The strategy which has been followed in the ILC design is to have redundant 
beam-based measurements of the incoming beam energy, capable of achieving a 
$10^{-4}$ relative precision on a single beam. This would be available in real 
time as a diagnostic tool to the operators and would provide the basis for the 
$\left<\sqrt{s}\right>$ determination. 
Additional physics reference channels, such as $e^+e^-\to\mu^+\mu^-\gamma$ where 
the muons are resonant with the known $Z$-mass, are then foreseen to provide 
valuable cross-checks of the collision scale, but only long after the data has 
been recorded.

The two primary methods planned for making precise beam energy measurements are a 
non-invasive BPM-based spectrometer, located upstream of the interaction point just 
after the energy collimators, and a synchrotron imaging detector which must be located 
downstream of the IP in the extraction line to the beam dump. The BPM-based device is 
modeled after the spectrometer built for LEP-II which was used to calibrate the energy 
scale for the $W$-mass measurement, although the parameters of the ILC version are tightly 
constrained by allowances on emittance dilution in the beam delivery system. 
The synchrotron imaging detector is similar in design to the spectrometer used at SLAC 
for the SLC program.  Both are designed to provide an absolute measurement of the beam
energy scale to a relative accuracy of $10^{-4}$. The downstream spectrometer, which 
observes the disrupted beam after collisions, can also measure the energy spectrum of 
the disrupted beam.

\subsection{Upstream Energy Spectrometer}
The canonical method to measure the beam energy $E_b$ upstream of the $e^+ e^-$--IP 
is the BPM-based spectrometer. A prototype test setup for such an instrument was 
proposed and commissioned in 2005, the T-474 experiment in the End Station A beamline 
at SLAC. The setup involves four dipole magnets and high-precision BPMs in front, 
behind and in between the magnets. 
ESA test beams operate at 10~Hz parasitically to PEP-II operation, with a bunch 
charge of $1.6 \cdot 10^{10}$ electrons, a bunch length of 500~$\mu$m and an energy 
spread of 0.15\%, i.e. with properties similar to ILC expectations. The beam energy 
is directly deduced from the offset measurements of 5~mm, which is also designed for 
the present ILC baseline energy spectrometer. When combining all the BPM stations to 
measure the precision of the orbit over the whole ESA-chicane beamline, a resolution 
of 0.82~$\mu$m in $x$ and 1.19~$\mu$m in $y$ was achieved. The system turned out to 
be stable at the micron level over the course of one hour. The long term stability 
was affected by relative scale drifts across all the BPMs. In particular, drifts of 
$\pm$10 $\mu$m were observed over 18~hours of operation. 
However, the stability of new designed ILC prototype BPMs located in the mid-chicane 
were stable to $\pm\;$0.25~$\mu$m over a period of one hour and $\pm$1 $\mu$m over 
the period of 18~hours. Their stability was influenced by low amplitude effects and 
mechanical vibration on short time scales. First results of the T-474 experiment were 
published, see e.g.~\cite{bib:ESA-BPM}, and support the successful operation of the 
testbench. Analyzing the data from 2007 runs is ongoing and final results are expected 
within the next few months. 
The T-474 experiment is not continuing past 2007 because of the 
cessation of the ESA test beam programs at SLAC.

\subsection{Extraction-Line Energy Spectrometer}
At the SLC, the WISRD (Wire Imaging Synchrotron Radiation Detector)~\cite{bib:WISRD} was 
used to measure the distance between two synchrotron stripes created by vertical bend 
magnets which surrounded a precisely-measured dipole that provided a horizontal bend 
proportional to the beam energy.  
This device achieved a precision of $\delta E_b/E_b \sim 2\times 10^{-4}$, where the 
limiting systematic errors were due to relative component alignment and magnetic field 
mapping. The ILC Extraction-Line Spectrometer (XLS) design is largely motivated by the 
WISRD experience.

The analyzing dipole for the XLS is provided by a vertical chicane just after the capture 
quad section of the extraction line, about 55 meters downstream of the interaction point. 
The chicane provides a $\pm 2$~mrad vertical bend to the beam and in both legs of the 
chicane horizontal wiggler magnets are used to produce the synchrotron light needed to 
measure the beam trajectory. The optics in the extraction line are designed to produce 
a secondary focus about 150 meters downstream of the IP, which coincides with the center 
of the polarimeter chicane and the Compton interaction point.  The synchrotron light 
produced by the wigglers will also come to a vertical focus at this point, and 
position-sensitive detectors in this plane arrayed outside the beampipe will measure 
the vertical separation between the synchrotron stripes.

With a total bend angle of 4~mrad, and a flight distance of nearly 100~meters, the 
synchrotron stripes will have a vertical separation of 400~mm, which must be measured 
to a precision of 40~$\mu$m to achieve the target accuracy of $10^{-4}$. In addition 
to the transverse separation of the synchrotron stripes, the integrated bending field 
of the analyzing dipole also needs to be measured and monitored to a comparable precision 
of $10^{-4}$. The distance from the analyzing chicane to the detectors needs to only be 
known to a modest accuracy of 1~cm.

In the original SLC WISRD, photoemission of electrons from thin wires on 100~$\mu$m
pitch was used as the detection mechanism.  This scheme suffered from several 
experimental issues, including cross-talk and RF pickup. For the XLS spectrometer, it 
has been proposed to use an array of radiation-hard 100~$\mu$m quartz fibers. These fibers 
do not detect the synchrotron light directly, but rather detect Cherenkov radiation from 
secondary electrons produced when the hard photons interact with material near the detector. 
At ILC beam energies, the critical energy for the synchrotron radiation produced in the 
XLS wigglers is several tens of MeV, well above the pair-production threshold, and copious 
numbers of relativistic electrons can be produced with a thin radiator in front of the 
fiber array.  The leading candidate for reading out these fibers are multi-anode PMTs 
from Hamamatsu, similar in design to those used in scintillating fiber calorimeters. 
The advantage of this scheme over wires is to produce a reliable, passive, rad-hard 
detector which does not suffer from cross-talk or RF pickup, and still allows for easy 
gain adjustment and a large dynamic range.

The energy spectrum of beam after collision contains a long, tail as a result of the 
beam-beam disruption in the collision process.  This disrupted beam spectrum is not a 
direct measure of the collision energy spectrum, but it is produced by the same physical 
process, and direct observation of this disrupted tail will serve as a useful diagnostic 
for the collision process. The position-sensitive detector in the XLS is designed to 
measure this beam energy spectrum down to 50\% of the nominal beam energy.  Near the peak, 
for a beam energy of $E_b = 250$~GeV, each 100 micron fiber spans an energy interval of 
125~MeV. Given a typical beam energy width of 0.15\%, this means the natural width of the 
beam energy will be distributed across at least a handful of fibers, which will allow 
the centroid to be determined with a precision better than the fiber pitch, and some 
information about the beam energy width can be extracted as well.

\subsection{R\&D on Alternative Methods}
R\&D on three alternative methods for precise beam energy measurements 
with 100~ppm accuracy is being carried out by different groups. 
The first method utilizes Compton backscattering, a magnetic spectrometer 
and precise position measurements of the electron beam, the centroid of 
the Compton photons and the kinematic edge of the Compton-scattered 
electrons~\cite{bib:EPWS-Viti, bib:EPWS-Muchnoi}. The spectrometer length 
needed is about 30~m and would be located near the upstream polarimeter. 
Precise position measurements approximately 25 meters downstream of an 
analysis magnet are needed with accuracies of  1~$\mu$m for the Compton 
photons, 10~$\mu$m for the Compton edge electrons and 0.5~$\mu$m for the 
beam electrons. Presently, a proposal to perform a proof-of-principle 
experiment at Novosibirsk is in preparation. Detailed studies are also 
in progress to understand whether a combination of the upstream 
polarisation chicane with a Compton energy spectrometer is possible.

The second method utilizes the synchrotron radiation (SR) emitted in the 
dipole magnets of the upstream BPM-based spectrometer~\cite{bib:SR-paper}. 
Accurate determination of the edges of the SR fan is needed. 
Studies include a direct measurement of the SR fan as well as the use 
of mirrors to deflect soft SR-light to detectors located away from the 
beamline.  Novel high-spatial resolution detectors are considered, and 
a gas amplification detector is now under study in Dubna with first 
results expected later this year.

A third method relies on the Resonance Absorption 
method~\cite{bib:EPWS-Melikian, bib:EPWS-Ghalumyan}. Under certain 
conditions, laser light can be absorbed by beam particles when both 
co-propagate in close proximity in a solenoid. The beam energy can be 
inferred from the measured dependence of light absorption on the magnetic 
field and laser wavelength. Studies are underway at Yerevan regarding 
theoretical uncertainties, and design of the laser system and laser 
light detectors.

\section{{\boldmath$Z$}-pole Calibration Data}
Precise energy and polarisation measurements at the $Z$-pole are not required in 
the RDR or in the ILC baseline parameters document. But, both measurements are 
important and should be included in the ILC baseline for the following reasons:
\begin{itemize}
\item  Polarimeter calibration and cross-check against physics based 
  polarisation measurements using the Blondel scheme;
\item  Calibration of energy measurement against $Z$-pole mass 
  (included in the RDR and in the ILC baseline parameters document) 
\item  Data from these calibration runs can also provide 
  significant statistics for physics measurements.
\end{itemize}
For a more detailed summary as to what can be done with the $Z$-pole calibration 
data the reader is referred to~\cite{bib:Zpole-Gudi}. \\
{\it\dred Recommendation:
  Include precise polarisation and energy measurements for $Z$-pole 
  calibration runs in the baseline. This is needed for calibration 
  cross checks of the polarimeters and energy spectrometers. 
}

\section{Upgrade to {\boldmath $\sqrt{s} = 1$}~TeV}
An energy upgrade to 1~TeV center-of-mass after the completion of the baseline 
programme should not be compromised in any way.
Measures should be taken not to render polarimetry impossible at beam energies 
higher than 250~GeV. This includes building beam diagnostics, especially the 
polarimeter chicanes, in a way that permits an easy upgrade to operate at high 
beam energies.

In this context again the combination of emittance diagnostics, machine protection 
and polarimetry as proposed in the RDR is extremely problematic, if not unfeasable, 
and we strongly recommend to separate these functions.
First of all, the ``scaled field'' or ``fixed dispersion'' operating scenario for 
the upstream chicane cannot be retained up to 500~GeV beam energy. This would lead 
to completely unacceptable synchrotron radiation losses and emittance blow-up. 
Secondly, if the dispersion would inevitably be scaled down to about 10~mm, the 
energy collimator will end up having a $\pm 1$~mm aperture (2~mm opening for 
a length of 3~m). 
This would be very problematic to operate even under nominal machine conditions 
and generate totally unacceptable background conditions for polarimetry.
Lastly, the performance of the laser-wire photon detector is already not really 
acceptable at lower beam energies, but at 500 GeV beam energy the proposed system 
will become unfeasible. 

\section{Conclusions and Recommendations}
The ``Workshop on Polarisation and Beam Energy Measurements at the ILC'' was 
accompanied by W.~Lorenzon (Univ. of Michigan) and K.~M\"onig (DESY-Zeuthen) 
as referees. In his conclusions, W.~Lorenzon stated that it is already all but 
trivial to provide or even prove an analyzing power precision at the 1\% level. 
He impressively showed this by discussing the setup and results of the 
``Spin Dance'' experiment performed at JLab (Thomas Jefferson National 
Accelerator Facility) in July 2000~\cite{bib:EPWS-Lorenzon-Concl, bib:SpinDance}.
In this experiment, a cross-normalisation of the relative analyzing power 
of the five electron polarimeters was performed to reveal possible 
systematic differences between the polarimeters that had not yet been 
accounted for. Although the systematic uncertainties of all polarimeters 
(1~Mott, 3~M\o{}ller, 1~Compton) were each evaluated individually, 
the experiment clearly showed significant discrepancies between the 
polarimeter results, even if the systematic uncertainties were included.

Furthermore, both referees argued that it is absolutely crucial to employ 
multiple devices for testing and controling the systematic uncertainties of 
each polarimeter~\cite{bib:PolScheme-Moenig, bib:EPWS-Lorenzon-Concl}. 
They also suggested to treat the upstream and downstream polarimeters as 
independent experiments and thus optimise them separately. This also implies 
that there is absolutely no need for both polarimeters to use the same type 
of laser, since the requirements and backgrounds also differ significantly. 
Their clear recommendations are to avoid any distraction from the ambitious 
goal of achieving a 0.25\% measurement of the beam polarisation. 

It was also strongly recommended to keep the initial positron polarisation 
of 30-45\% to improve the gain in effective luminosity and enable physics 
measurements, which would not be possible without positron polarisation. 
In any case the positron beam will be polarised from the start due to the 
helical undulator source used to generate the positrons.

The motivation for having both, upstream and downstream diagnostics, includes 
complementarity, redundancy and intercalibration of the systems. As can easily 
be seen from the experiences at SLC and at LEP, independent measurements 
proved to be important for both, polarimetry and energy measurements. Also, 
over a decade of operational experience with multiple Compton polarimeters 
at HERA clearly demonstrated the necessity for such redundancy, both in terms 
of systematic cross checks and in terms of operational reliability.

\section*{Acknowledgments}
The authors would like to thank the experiments support group of DESY, Zeuthen 
for the well-prepared workshop, the exceptionally good working atmosphere and 
their general hospitality, including coffee, tea, cake and cookies at odd times.


\begin{footnotesize}


\end{footnotesize}

 
\end{document}